\documentclass[aps,prl,showpacs,superscriptaddress,floatfix, twocolumn]{revtex4}
\usepackage{graphicx}

\newcommand{\xv}{\vec{x}}
\newcommand{\kv}{\vec{k}}
\newcommand{\pv}{\vec{p}}

\newcommand{\be}{\begin{equation}}
\newcommand{\ee}{\end{equation}}
\newcommand{\bea}{\begin{eqnarray}}
\newcommand{\eea}{\end{eqnarray}}

\begin{document}

\title{Constant flux relation for driven dissipative systems}
\author{Colm Connaughton}
\email{connaughtonc@gmail.com} 
\affiliation {Center for Nonlinear Studies,
Los Alamos National Laboratory, Los Alamos, NM 87545, USA}
\author{R. Rajesh}
\email{rrajesh@imsc.res.in}
\affiliation{
Institute of Mathematical Sciences, CIT Campus, Taramani, Chennai-600013,
India}
\author{Oleg Zaboronski}
\email{olegz@maths.warwick.ac.uk}
\affiliation{Mathematics Institute, University of Warwick, Gibbet Hill
Road, Coventry CV4 7AL, UK}

\date{\today}

\begin{abstract}
Conservation laws constrain the stationary state statistics
of driven dissipative systems because the average flux of a
conserved quantity between driving and dissipation scales should be
constant. This requirement leads to a universal scaling law for
flux-measuring correlation functions, which generalizes the 4/5-th law of
Navier-Stokes turbulence. We demonstrate the utility of this simple idea
by deriving new exact scaling relations for models of aggregating particle
systems in the fluctuation-dominated regime and for energy and wave action
cascades in models of strong wave turbulence.
\end{abstract}

\maketitle

A major challenge of theoretical physics is to develop a general
formalism for understanding properties of systems far from
equilibrium. In the absence of such a formalism, exact results play
an important role. They act as precursors for more general
theories and as checkpoints for phenomenological theories and
numerical methods. In this paper we derive an exact relation
satisfied by a particular correlation function for an important
class of {\it turbulent} non-equilibrium systems.

By turbulent systems, we understand not just fluid turbulence, but
a more general class of driven dissipative systems
characterised by the presence of an inertial range. This is a 
wide range of scales separating the driving and dissipation. In the inertial 
range, dynamics is expected to be
universal, i. e. independent of details of driving and
dissipation. In the steady state, the inertial range dynamics is
characterised by constant fluxes of conserved quantities
from driving to dissipation scale. A well known example is
Navier-Stokes (NS) turbulence \cite{frischBook}. Energy
injected into the system at a large length scale $L$  is
dissipated at a length scale $l\sim \nu^{3/4}$, where $\nu$ is the
kinematic viscosity. In the inertial range $l \ll r \ll L$, there
is a constant energy flux directed to small scales.

If the inertial range dynamics is scale invariant, the
constant flux of a conserved quantity always fixes the scaling of a certain 
correlation function. An example
is Kolmogorov's $4/5$ law of 3 dimensional NS
turbulence \cite{K41b} which states that, in the
inertial range,
\be
\langle \big[ v_l(\vec{r},t)-v_l(\vec{0},t) \big]^3 \rangle
= -\frac{4}{5}\epsilon r, \quad l \ll r \ll L, \label{4/5th}
\ee
where $v_l(\vec{r},t)$ is the longitudinal component of the velocity at
$\vec{r}$ at time $t$ and $\epsilon$ is the average energy flux. Pedagogical
derivations of this result are in \cite{landauBook, frischBook}.
Other examples are the Monin-Yaglom relation for passive
advection (see \cite{Grisha1} and references therein) and Burgers
turbulence where
integrals of motion can be used to prove extreme
anomalous scaling of all structure functions \cite{Grisha2}.

A general law relating scaling properties of correlation functions
to inertial range conservation laws is still missing. As a result, no 
counterpart of the $4/5$-law is
known for such important turbulent systems such as $3-$
and $4$-wave turbulence and cluster-cluster aggregation. In this
paper we derive  a general constant flux relation (CFR) which
should be obeyed  by all flux-determined correlations of turbulent
systems. We start with a heuristic argument, followed by
an analytic application of this argument to the following specific models:

 \begin{itemize}
\item Mass model (MM): Consider particles of positive mass on a $d$-dimensional 
lattice. They evolve in time by diffusion, at rate $D$, aggregation of
particles at the same site, at rate $\lambda (m_1,m_2)$, and injection of
particles of mass $m_0$ at rate $J/m_0$. For more detail, see \cite{CRZtmlong}.

\item Charge model (CM): The model is similar to MM except for the
following modifications. The masses (now charges) can be positive
or negative. Particles of charge $\pm m_0$ are input at the same
rate $J/m_0^2$.

\item  Wave turbulence (WT): An ensemble of interacting dispersive
waves maintained far from equilibrium by interaction with external 
sources and sinks. For details and a review see \cite{zakharovBook, newell2001}.
\end{itemize}

Together, these models relate to a broad range of disciplines,  far
beyond the common applications of fluid turbulence. Applications of the MM
or CM include
submonolayer epitaxial thin film growth \cite{KMR1999}, river
networks \cite{scheidegger1967,dodds1999}, force fluctuations in
granular bead packs \cite{coppersmith1996} and non-equilibrium
phase transitions \cite{MKB2000,rajesh2004}. The constant kernel MM maps 
\cite{dhar1999} onto the directed abelian sandpile, one of the 
first models generating power laws from simple dynamical rules \cite{dhar1989}.
WT is realized in such diverse situations as
capillary and gravity waves in liquids, electromagnetic waves in
non-linear optics and Alfven waves in plasmas (see
\cite{zakharovBook,zakharov2004} for more applications).

We present a heuristic argument for a general turbulent system. It is
necessarily abstract, hiding much of the physics of specific systems. 
We believe that the general approach only acquires clarity after
application to concrete examples. The idea is very simple. First identify
the conserved quantity expected to cascade, we shall call it $I$,  and the 
space in which it flows.
Second, use the equation of motion to write down a Boltzmann-like 
continuity equation for the average density of $I$ in this space.
This equation identifies a flux-carrying correlation function, $\Pi$, and
a nonlinear coupling, $T$, controlling the redistribution of $I$.
Third, use dimensional analysis to find the scaling of
$\Pi$ corresponding to constant flux.
Suppose the flux of $I$ occurs in a $d$-dimensional space,
spanned by $\pv$. The degrees of freedom are at least partially indexed by 
$\vec{p}$. We denote them by
$u_{\pv}$. For example, in three dimensional NS turbulence, energy
transfer occurs between all modes in Fourier space, so $d=3$. Also
$\pv = \vec{k}$, the wave number, and $u_{\pv}$ is $\vec{v}
(\vec{k})$, the velocity field. For surface waves,
motion occurs in three dimensions but energy is transferred between
wave trains indexed by a two dimensional wave-vector, so $d=2$. For
particle aggregation problems, mass is transferred in one dimensional mass
space so $d=1$ even though the particles may be diffusing on a
lattice of arbitrary spatial dimension.

The system must have an inertial range in $\pv$-space, within which the
dynamics is entirely dominated by the $I$-conserving interactions. Denote
the flux by $\vec{J}(\pv)$. Assume for 
simplicity, that the inertial range dynamics is isotropic, so
we can average over angles in $\pv$-space.  Let $\rho_I$ be 
the density, and $J(p)$ the flux, of $I$ in $p$-space, 
where 
$p=\left|\pv \right|$. In the inertial range the flux of $I$ through a
sphere of radius $p$ does not depend on $p$. Therefore,
$\vec{J}(\pv)=C ~ p^{-d} \pv$. On the other hand, from the equation of
motion, one can always derive a continuity equation for $I$ in $p$-space taking
the general form:
\be
\label{eq-heuristicContEq}
\partial_p J(p)= -\dot{\rho}_{I}=\int
\prod_{m=1}^{n-1}d^dp_m T(p,p_1,p_2, \ldots, p_{n-1})\Pi.
\ee
$T$ is a non-linear coupling and $\Pi$ is a
correlation function of the $u_{\pv}$, whose precise forms must
be calculated for each specific example. In   
Eq.~(\ref{eq-heuristicContEq}) it is assumed that the non-linearity is of
degree $n-1$ in the $u_{\pv}$. We take $T$ to be homogeneous of degree $\beta$:
$T(\Lambda p) = \Lambda^\beta T(p)$.  
Equating scaling exponents, we find
\bea
\Pi (p)\sim p^{-nd-\beta}.
\label{cfr}
\eea

How does this relate to Kolmogorov's $4/5$ law? Eq.~(\ref{4/5th}) is
usually derived in real space although the flux is in $\kv$-space, so
the correspondence is not clear. A $\kv$-space derivation of the $4/5$ law,
along exactly the lines outlined here is indeed possible. It can
be found in \cite{kuznetsov1981}. Even earlier, a similar approach was taken by 
Kraichnan in his seminal work on 2D turbulence \cite{67Kra}. 
To summarise, for NS, $n=3$ (nonlinearity 
is quadratic), $\beta=1$ (nonlinearity contains one derivative)
and $\Pi \sim \langle v_l(\kv_1)v_l(\kv_2) v_l(\kv_3) \rangle$.
By  Eq.~(\ref{cfr}), $\Pi  \sim
k^{-1-3d}$. Thus $\langle (v_l(r)-v_l(0))^3 \rangle \sim
r^{1+3d-3d}=r$, consistent with Eq.~(\ref{4/5th}).

The remainder of this letter is dedicated to the illustration and 
analytical verification Eq.~(\ref{cfr}) for the specific models listed above.
We illustrate the practical meaning of the steps outlined above by explicitly
identifying $\Pi$ and $T$ and, by exact computations, show that
Eq.~(\ref{cfr}) is more than dimensional analysis. The examples extend 
the applicability of the $4/5$-law beyond the usual cases. MM is a pedagogical 
example of flux conservation outside of hydrodynamics. CM illustrates that the 
idea is also relevant  to  statistical conservation laws. WT demonstrates
the approach for non-quadratic invariants.

First, consider the MM. The conserved quantity is mass. The mass flux is
in mass space. Let the reaction rate
$\lambda_{m_1,m_2}$ be a homogeneous function of degree $\beta$.
Let $N(m,\xv,t)$ be the local mass distribution. The continuity
equation for the average mass density, $m\langle
N(m,\xv,t)\rangle$ takes the following form in the limit
$m_0\rightarrow 0, ~t \rightarrow \infty$:
 \be \label{hopfTM}
m \int_0^{\infty}\!\!\! dm_{1} dm_2  \left[ I_{0;1,2} -
I_{2;1,0} - I_{1;0,2} \right]=0,
 \ee
where $I_{0;1,2}= \lambda_{m_1,m_2}  C(m_1,\!m_2)
\delta(m\!-m_{1}-m_{2})$ and $C(m_{1}, m_{2},t)=\langle
N(m_{1},\xv,t)N(m_{2},\xv,t)-\frac{1}{\Delta
x^{d}}\delta(m_{1}-m_{2})N(m_{1},\xv,t) \rangle$ is the average
number of pairs of particles of masses $m_1$ and $m_2$ per lattice
site. $\Delta x$ is the lattice spacing. Eq.~(\ref{hopfTM}) is not supposed to
be obvious. A detailed derivation for the case $\beta=0$ can be found in 
\cite{CRZtmlong}.
Let us look for homogeneous
solutions to Eq.~(\ref{hopfTM}) of degree $h$, $i. e.$,
$C(\Lambda m_{1}, \Lambda m_{2}) = \Lambda^h C(m_{1},
m_{2})$. The value of $h$ is determined by the balance between the
'plus' and 'minus' terms in the left hand side of
Eq.~(\ref{hopfTM}). This balance can be made explicit by applying the
following Zakharov Transformations \cite{zakharovBook, CRZtmshort} to the 
'minus' terms:
\begin{eqnarray}
\mbox{Second integral:} & &\big( m_{1}, m_{2}\big) \rightarrow \left(
\frac{m m_{1}}{m_{2}},\frac{m^2}{m_{2}}\right), \\
\mbox{Third integral:} & & \big( m_{1}, m_{2}\big) \rightarrow \left(
\frac{m^2}{m_{1}},\frac{m m_2}{m_{1}} \right), \label{ztTM}
\end{eqnarray}
after which we obtain
 \be 0=\int_{0}^{\infty} \int_0^{\infty} dm_{1} dm_2
 I_{m;m_1,m_2} \left[ m^{y} -m_1^{y} -m_2^{y} \right] \label{cfTM}
 \ee
 where $y=-h-\beta-3$. 
Eq.~(\ref{cfTM}) is satisfied only if $y=1$: 
the function $f(m_1, m_2)=(m_1+m_2)^y-m_1^y-m_2^y$ is sign
definite for any $y \neq1$ whereas $I_{m,m_1,m_2}$ is
non-negative. Therefore the integral in the right hand side of
Eq.~(\ref{cfTM}) is zero only for $y=1$. Consequently,
$h=-3-\beta$, 
is the unique scaling solution, under an extra assumption of
convergence of collision integrals, which will be discussed
elsewhere. For $\beta=0$, the scaling $h=-3-\beta$ is  already 
established. It is trivially true for $d>2$ where mean field holds,
in $d=2$ due to cancellation of logarithmic corrections
\cite{CRZtmlong} and in $d<2$ by an exact solution
\cite{RM2000}.

The scaling of the flux correlation function $\Pi=m C \delta(m) \sim
m^{-3-\beta}$ agrees with general CFR, Eq.~(\ref{cfr}), for $d=1$,
$n=3$. Note that the result, remarkably, does not depend on the transport
properties of the particles such as mass dependence of the diffusion
coefficient. 

Next, consider CM. There are two conserved quantities: average charge
and average squared charge. Charge is conserved in elementary collisions, but 
squared charge is conserved only on
average. To see this, consider the coagulations between the
following pairs of charges: $(m_{1}, m_{2})$, $(-m_{1}, m_{2})$,
$(m_{1}, -m_{2})$ and $(-m_{1}, -m_{2})$. They occur with equal
probability and the total squared charge in the initial and final states are
equal. Therefore, squared charge is
conserved {\it on average}. Total flux of charge is zero by the symmetry of 
input. Therefore, for the CM, CFR is
associated with a flux of squared charge. The corresponding
continuity equation is
 \be \label{hopfCM} m^2 \int_0^{\infty}\!\!\!\!\!
dm_{1} dm_2 \left[ I'_{0;1,2}\! -\! I'_{2;1,0}\! -\!
I'_{1;0,2} \right]=0, \ee
 where $I'_{0;1,2}\!=\!
K_{m_1,m_2} C(m_1,\!m_2) \delta[m \pm m_1\pm m_2]$. Here
$C(m_1,m_2)$ has the same meaning as in the MM. Due to symmetry
$C(-m_1,m_2)=C(m_1,m_2)$. We assume $K_{m_1,m_2}$ to be
homogeneous of degree $\beta$. Also, $\delta(x \pm y \pm
 z)=\sum_{m,n=0}^{1}\delta(x-(-1)^my-(-1)^nz)$. The following ZT is applied to Eq.~(\ref{hopfCM}):
\begin{eqnarray}
\mbox{First integral:} & &\left( m_{1}, m_{2}\right) \rightarrow \left(
\frac{m^2}{m_{1}},\frac{m m_2}{m_{1}}\right), \\
\mbox{Second integral:} & & \left( m_{1}, m_{2}\right) \rightarrow \left(
\frac{m^2}{m_{2}},\frac{m m_1}{m_{2}} \right), \label{ztCM}
\end{eqnarray}
and integration over $m_1$ is performed to obtain
 \begin{eqnarray}
&0&=\int_0^{\infty}dm_{2} K(m,m_2)C(m,m_2)  \bigg[ m^2 \left(
\frac{m}{|m-m_2|}\right)^{-y}\nonumber \\ &+&m^2\left(
\frac{m}{|m+m_2|}\right)^{-y}-2m^2-2m^2\left(\frac{m}{m_2}
\right)^{-y} \bigg], \label{cfCM}
 \end{eqnarray}
 where $y=-h-\beta-2$ and $h$ is the homogeneity exponent of $C(m_1,m_2)$ to be determined. If $y=2$, then
Eq.~(\ref{cfCM}) is satisfied as the integrand vanishes
identically due to square charge conservation. Generalizing the
argument made in case of MM, it is possible to check that $y=2$ is
the unique scaling solution, given locality. Thus,
$h=-4-\beta.$
The scaling of the flux measuring correlation function $\Pi=m^2
C(m)\delta(m)\sim m^{-3-\beta}$ is consistent with Eq.~(\ref{cfr}) for $d=1$ 
and $n=3$.

Finally we consider wave turbulence. It is a Hamiltonian system 
with canonical variables
$\{a_{\vec{k}}, \bar{a}_{\vec{k}}\}_{\vec{k}\in {\bf R}^d}$:
\be H=\int d\vec{k} [~\omega(k)
\bar{a}_{\vec{k}}a_{\vec{k}}+ u(\vec{k})~]. \label{ham4} 
\ee 
$\omega$ is a  homogeneous function of $\vec{k}$ and $u(\vec{k})$
is the nonlinear part of the energy density. When $u=0$, the
Hamiltonian describes free waves with dispersion law
$\omega=\omega(k)$, $\vec{k}$ being the wave vector. The
equations of motion are 
\be 
\label{eq-WTeqnMotion}
\dot{a}_{\vec{k}}= i
\left[ \omega(k) a_{\vec{k}} + \frac{\delta U}{\delta
\bar{a}_{\vec{k}}}\right], 
\ee 
where $U=\int d\vec{k} u(\vec{k})$. The equation of motion
conserves $H$. There may be additional conservation
laws depending on the form of $u(\vec{k})$. The system 
is maintained far from equilibrium by adding appropriate
forcing and dissipation terms to Eq.~(\ref{eq-WTeqnMotion}).
Most commonly, $u(\vec{k})$ has degree 3 or 4 in $a_{\vec{k}}$.

We first study the energy cascade
which conserves energy in the inertial range. As a result,
the average flux of energy is constant, which leads to the
continuity equation
 \be \left\langle \dot{u} (\vec{k}) -
\dot{\bar{a}}_{\vec{k}} \frac{\delta U}{\delta \bar{a}_{\vec{k}}} -
\dot{a}_{\vec{k}} \frac{\delta U}{\delta a_{\vec{k}}}
\right\rangle=0, \label{eq:WT}
 \ee
 where $U$ is the interaction part of the Hamiltonian. We now consider two cases:
3-wave turbulence  (n=3) where
 \be u =\!\! \int \!\!\prod_{i=1}^{2}
d\vec{k}_i \delta (\vec{k}\!-\!\vec{k}_1\!-\!\vec{k}_2)
T_{k;k_1,k_2} [ \bar{a}_{\vec{k}} a_{\vec{k}_1} a_{\vec{k}_2}+c.c],
\label{eq:3WT}
 \ee
 and 4-wave turbulence (n=4)  where \bea u &=& \int
 \prod_{i=1}^{3} d\vec{k}_i \delta
 (\vec{k}+\vec{k}_1-\vec{k}_2-\vec{k}_3)
 T_{k,k_1;k_2,k_3} \nonumber\\
 && \times
[ \bar{a}_{\vec{k}} \bar{a}_{\vec{k}_1} a_{\vec{k}_2}
a_{\vec{k}_3}+c.c.]. \label{eq:4WT} \eea
In both cases, we assume $T$ to be homogeneous of 
degree $\beta$. For 3-wave turbulence Eq.~(\ref{eq:WT}), reduces to
\be 
\int
\prod_{i=1}^2 (dk_i k_i^{d-1}) \big[T_{k;k_1,k_2} \Pi_{0;1,2}
-T_{k_1;k,k_2} \Pi_{1;0,2}\big]=0,
\ee
where $\Pi_{0;1,2} = \int \prod_{i=0}^{2}  d\Omega_i  \langle
Re( a_{\vec{k}} \dot{\bar{a}}_{\vec{k}_1} \bar{a}_{\vec{k}_2} )
\rangle$ is the flux correlation function. Note an essential difference
between the flux correlation function for total energy  and that for
quadratic energy. Total energy flux depends on  correlations between fields and 
time derivatives of fields whereas quadratic energy flux would depend on
correlations between fields only. Transforming
$(k_{1},k_{2})\rightarrow (k^2/k_{1},k k_{2}/k_{1})$ in the second
integral, we obtain
 \be 
\label{eq:3WCH}
0=\int \prod_{i=1}^2 (dk_i k_i^{d-1})
T_{k;k_1,k_2} \Pi_{0;1,2} \left[k^{y}-k_1^{y}\right],
 \ee
where $y=-3 d - h - \beta$ and $h$ is the degree of homogeneity of
$\Pi$. Therefore, $y=0$, or
$h= - 3 d -\beta$
 solves Eq.(\ref{eq:3WCH}). The result $\Pi
\sim k^{-3d-\beta}$ agrees with  Eq.~(\ref{cfr}), as the
order of nonlinearity is $n=3$, dimensionality of flux space is
$d$ and homogeneity index of interaction kernel $T$ is $\beta$. It
is also consistent with Kolmogorov-Zakharov (KZ) theory describing
turbulence of weakly non-linear waves \cite{zakharovBook}: in this
limit the wave action density corresponding to an energy cascade is
$n(k)\sim k^{-(\beta+d)}$. Within KZ theory, $\Pi \sim \omega
Im \langle a \bar{a}^2 \rangle \sim \omega T \delta (\omega)
\delta^d (k) n(k)^2 \sim k^{-3d-\beta}$. We stress however
that, unlike KZ theory, this result is equally valid for
strong wave turbulence.

A similar analysis for the 4 wave case yields $h=-4 d -\beta$ for the 
homogeneity degree of the correlation function $\Pi_{0,1;2,3} =
\int \prod_{i=0}^{3}  d\Omega_i \langle Re( \dot{a}_{\vec{k}}
a_{\vec{k}_1} \bar{a}_{\vec{k}_2} \bar{a}_{\vec{k}_3}) \rangle$. This
is also consistent with KZ theory for weak non-linearity
and with Eq.~(\ref{cfr}) for the order of non-linearity $n=4$.

In addition to energy,  4-wave turbulence also
conserves wave action $N=\int d\vec{k} n(\vec{k})$, where $\langle
\bar{a}_{\vec{k}_1} a_{\vec{k}_2} \rangle = n(\vec{k}_1)
\delta(\vec{k}_1-\vec{k}_2)$. There is a CFR
associated with the
cascade of wave action. The flux correlation function is
$\Pi_{0,1;2,3} = \int \prod_{i=0}^{3} d \Omega_i Im\langle
\bar{a}_{\vec{k}} \bar{a}_{\vec{k}_1} a_{\vec{k}_2} a_{\vec{k}_3})
\rangle.$ Assume that $\Pi$ is a scaling function of degree $h$.
Applying ZT to the continuity equation we find
This is consistent with Eq.~(\ref{cfr}), for order
of non-linearity $n=4$, dimensionality of flux space $d$ and
homogeneity degree of interaction kernel $\beta$. In the limit of
small non-linearity, the scaling $h=-4d-\beta$ can be also derived from KZ
theory: for weakly non-linear waves, wave action density in the
wave action cascade is $n(k) \sim k^{-(2 \beta-\alpha+3d)/3}$, where
$\alpha$ is dispersion exponent. On the other hand, within KZ
theory, $\Pi \sim T \delta^d(k)\delta(\omega) n(k)^3\sim
k^{-4d-\beta}$.

This ends our illustration of specific examples where the CFR idea, 
expressed heuristically in Eq. (\ref{cfr}), is used to derive new scaling
laws respected even by strongly interacting systems. Probably 
there are other examples waiting to be analysed using the ideas presented
in this letter. In closing we should emphasise  that, although the results for
specific systems analysed here are exact, the CFR is generally not a theorem.
Application of the ZT implicitly assumes convergence of the collision integral
on the scaling solution for the flux carrying correlation function (in the wave
turbulence literature, this is referred to as the condition of locality). This 
assumption is usually not readily verifiable, even a posteriori, and will 
require considerable additional effort both theoretically and numerically. 
Nevertheless, there are specific cases where these issues are simplified
which we would like to mention. For the MM, the constant kernel case 
($\beta=0$), can be reduced to a differential equation which can be explicitly
solved. For wave turbulence with certain types of ultra-local interaction
coefficients, CFR can be derived as a theorem by solving some simple
recursion relations. It is not our intention to present more detailed analysis 
of these particulars here but rather to give an assurance that the assumption
of locality has a basis in reality.

To summarize, we demonstrated, using a number of examples, that
the scaling of a particular correlation function can be determined
when there is a constant flux of a physical quantity in
scale-invariant turbulent dynamics. This provides an extension
of Kolmogorov's 1941 work on NS turbulence and Yaglom's
work on passive advection to a diverse class of systems.

\noindent{\bf Acknowlegements}\\
We would like to thank G. Falkovich and I. Kolokolov for
interesting discussions and useful suggestions. This work was partially carried 
out under the auspices of the National Nuclear Security Administration of the 
U.S. Department of Energy at Los Alamos National Laboratory under Contract No. 
DE-AC52-06NA25396.


\begin{thebibliography}{21}
\expandafter\ifx\csname natexlab\endcsname\relax\def\natexlab#1{#1}\fi
\expandafter\ifx\csname bibnamefont\endcsname\relax
  \def\bibnamefont#1{#1}\fi
\expandafter\ifx\csname bibfnamefont\endcsname\relax
  \def\bibfnamefont#1{#1}\fi
\expandafter\ifx\csname citenamefont\endcsname\relax
  \def\citenamefont#1{#1}\fi
\expandafter\ifx\csname url\endcsname\relax
  \def\url#1{\texttt{#1}}\fi
\expandafter\ifx\csname urlprefix\endcsname\relax\def\urlprefix{URL }\fi
\providecommand{\bibinfo}[2]{#2}
\providecommand{\eprint}[2][]{\url{#2}}

\bibitem[{\citenamefont{Frisch}(1995)}]{frischBook}
\bibinfo{author}{\bibfnamefont{U.}~\bibnamefont{Frisch}},
  \emph{\bibinfo{title}{Turbulence: The Legacy of A. N. Kolmogorov}}
  (\bibinfo{publisher}{Cambridge University Press},
  \bibinfo{address}{Cambridge}, \bibinfo{year}{1995}).

\bibitem[{\citenamefont{Kolmogorov}(1941)}]{K41b}
\bibinfo{author}{\bibfnamefont{A.~N.} \bibnamefont{Kolmogorov}},
  \bibinfo{journal}{Dokl. Akad. Nauk. SSSR} \textbf{\bibinfo{volume}{32}},
  \bibinfo{pages}{16} (\bibinfo{year}{1941}), \bibinfo{note}{reprinted in Proc.
  Roy. Soc. Lond. A, 434, 15-17, (1991)}.

\bibitem[{\citenamefont{Landau and Lifshitz}(1987)}]{landauBook}
\bibinfo{author}{\bibfnamefont{L.~D.} \bibnamefont{Landau}} \bibnamefont{and}
  \bibinfo{author}{\bibfnamefont{E.~M.} \bibnamefont{Lifshitz}},
  \emph{\bibinfo{title}{Fluid Mechanics}} (\bibinfo{publisher}{Pergamon Press},
  \bibinfo{address}{London}, \bibinfo{year}{1987}), \bibinfo{edition}{2nd} ed.

\bibitem[{\citenamefont{Falkovich et~al.}(2001)\citenamefont{Falkovich,
  Gawedzki, and Vergassola}}]{Grisha1}
\bibinfo{author}{\bibfnamefont{G.}~\bibnamefont{Falkovich}},
  \bibinfo{author}{\bibfnamefont{K.}~\bibnamefont{Gawedzki}}, \bibnamefont{and}
  \bibinfo{author}{\bibfnamefont{M.}~\bibnamefont{Vergassola}},
  \bibinfo{journal}{Rev. Modern Phys.} \textbf{\bibinfo{volume}{73}},
  \bibinfo{pages}{913} (\bibinfo{year}{2001}).

\bibitem[{\citenamefont{Falkovich}(2004)}]{Grisha2}
\bibinfo{author}{\bibfnamefont{G.}~\bibnamefont{Falkovich}}, in
  \emph{\bibinfo{booktitle}{Encyclopedia of Nonlinear Science}}, edited by
  \bibinfo{editor}{\bibfnamefont{A.}~\bibnamefont{Scott}}
  (\bibinfo{publisher}{Routledge}, \bibinfo{address}{New York and London},
  \bibinfo{year}{2004}).

\bibitem[{\citenamefont{Connaughton et~al.}(2006)\citenamefont{Connaughton,
  Rajesh, and Zaboronski}}]{CRZtmlong}
\bibinfo{author}{\bibfnamefont{C.}~\bibnamefont{Connaughton}},
  \bibinfo{author}{\bibfnamefont{R.}~\bibnamefont{Rajesh}}, \bibnamefont{and}
  \bibinfo{author}{\bibfnamefont{O.}~\bibnamefont{Zaboronski}},
  \bibinfo{journal}{Physica D} \textbf{\bibinfo{volume}{222}},
  \bibinfo{pages}{97} (\bibinfo{year}{2006}).

\bibitem[{\citenamefont{Zakharov et~al.}(1992)\citenamefont{Zakharov, Lvov, and
  Falkovich}}]{zakharovBook}
\bibinfo{author}{\bibfnamefont{V.}~\bibnamefont{Zakharov}},
  \bibinfo{author}{\bibfnamefont{V.}~\bibnamefont{Lvov}}, \bibnamefont{and}
  \bibinfo{author}{\bibfnamefont{G.}~\bibnamefont{Falkovich}},
  \emph{\bibinfo{title}{Kolmogorov Spectra of Turbulence}}
  (\bibinfo{publisher}{Springer-Verlag}, \bibinfo{address}{Berlin},
  \bibinfo{year}{1992}).

\bibitem[{\citenamefont{Newell et~al.}(2001)\citenamefont{Newell, Nazarenko,
  and Biven}}]{newell2001}
\bibinfo{author}{\bibfnamefont{A.}~\bibnamefont{Newell}},
  \bibinfo{author}{\bibfnamefont{S.}~\bibnamefont{Nazarenko}},
  \bibnamefont{and} \bibinfo{author}{\bibfnamefont{L.}~\bibnamefont{Biven}},
  \bibinfo{journal}{Physica D} \textbf{\bibinfo{volume}{152-153}},
  \bibinfo{pages}{520} (\bibinfo{year}{2001}).

\bibitem[{\citenamefont{Krapivsky et~al.}(1999)\citenamefont{Krapivsky, Mendes,
  and Redner}}]{KMR1999}
\bibinfo{author}{\bibfnamefont{P.~L.} \bibnamefont{Krapivsky}},
  \bibinfo{author}{\bibfnamefont{J.~F.~F.} \bibnamefont{Mendes}},
  \bibnamefont{and} \bibinfo{author}{\bibfnamefont{S.}~\bibnamefont{Redner}},
  \bibinfo{journal}{Phys. Rev. B} \textbf{\bibinfo{volume}{59}},
  \bibinfo{pages}{15950} (\bibinfo{year}{1999}).

\bibitem[{\citenamefont{Scheidegger}(1967)}]{scheidegger1967}
\bibinfo{author}{\bibfnamefont{A.~E.} \bibnamefont{Scheidegger}},
  \bibinfo{journal}{Bull. I. A. S. H.} \textbf{\bibinfo{volume}{12}},
  \bibinfo{pages}{15} (\bibinfo{year}{1967}).

\bibitem[{\citenamefont{Dodds and Rothman}(1999)}]{dodds1999}
\bibinfo{author}{\bibfnamefont{P.~S.} \bibnamefont{Dodds}} \bibnamefont{and}
  \bibinfo{author}{\bibfnamefont{D.~H.} \bibnamefont{Rothman}},
  \bibinfo{journal}{Phys. Rev. E} \textbf{\bibinfo{volume}{59}},
  \bibinfo{pages}{4865} (\bibinfo{year}{1999}).

\bibitem[{\citenamefont{Coppersmith et~al.}(1996)\citenamefont{Coppersmith,
  Liu, Majumdar, Narayan, and Witten}}]{coppersmith1996}
\bibinfo{author}{\bibfnamefont{S.~N.} \bibnamefont{Coppersmith}},
  \bibinfo{author}{\bibfnamefont{C.}~\bibnamefont{Liu}},
  \bibinfo{author}{\bibfnamefont{S.}~\bibnamefont{Majumdar}},
  \bibinfo{author}{\bibfnamefont{O.}~\bibnamefont{Narayan}}, \bibnamefont{and}
  \bibinfo{author}{\bibfnamefont{T.~A.} \bibnamefont{Witten}},
  \bibinfo{journal}{Phys. Rev. E} \textbf{\bibinfo{volume}{53}},
  \bibinfo{pages}{4673} (\bibinfo{year}{1996}).

\bibitem[{\citenamefont{Majumdar et~al.}(2000)\citenamefont{Majumdar,
  Krishnamurthy, and Barma}}]{MKB2000}
\bibinfo{author}{\bibfnamefont{S.~N.} \bibnamefont{Majumdar}},
  \bibinfo{author}{\bibfnamefont{S.}~\bibnamefont{Krishnamurthy}},
  \bibnamefont{and} \bibinfo{author}{\bibfnamefont{M.}~\bibnamefont{Barma}},
  \bibinfo{journal}{Phys. Rev. E} \textbf{\bibinfo{volume}{61}},
  \bibinfo{pages}{6337} (\bibinfo{year}{2000}).

\bibitem[{\citenamefont{Rajesh}(2004)}]{rajesh2004}
\bibinfo{author}{\bibfnamefont{R.}~\bibnamefont{Rajesh}},
  \bibinfo{journal}{Phys. Rev. E} \textbf{\bibinfo{volume}{69}},
  \bibinfo{pages}{036128} (\bibinfo{year}{2004}).

\bibitem[{\citenamefont{Dhar}(1999)}]{dhar1999}
\bibinfo{author}{\bibfnamefont{D.}~\bibnamefont{Dhar}},
  \bibinfo{howpublished}{preprint arXiv:cond-mat/9909009}
  (\bibinfo{year}{1999}).

\bibitem[{\citenamefont{Dhar and Ramaswamy}(1989)}]{dhar1989}
\bibinfo{author}{\bibfnamefont{D.}~\bibnamefont{Dhar}} \bibnamefont{and}
  \bibinfo{author}{\bibfnamefont{R.}~\bibnamefont{Ramaswamy}},
  \bibinfo{journal}{Phys. Rev. Lett.} \textbf{\bibinfo{volume}{63}},
  \bibinfo{pages}{1659} (\bibinfo{year}{1989}).

\bibitem[{\citenamefont{Zakharov et~al.}(2004)\citenamefont{Zakharov, Dias, and
  Pushkarev}}]{zakharov2004}
\bibinfo{author}{\bibfnamefont{V.~S.} \bibnamefont{Zakharov}},
  \bibinfo{author}{\bibfnamefont{F.}~\bibnamefont{Dias}}, \bibnamefont{and}
  \bibinfo{author}{\bibfnamefont{A.}~\bibnamefont{Pushkarev}},
  \bibinfo{journal}{Phys. Rep.} \textbf{\bibinfo{volume}{398}},
  \bibinfo{pages}{1} (\bibinfo{year}{2004}).

\bibitem[{\citenamefont{Kuznetsov and Lvov}(1981)}]{kuznetsov1981}
\bibinfo{author}{\bibfnamefont{E.}~\bibnamefont{Kuznetsov}} \bibnamefont{and}
  \bibinfo{author}{\bibfnamefont{V.}~\bibnamefont{Lvov}},
  \bibinfo{journal}{Physica D} \textbf{\bibinfo{volume}{2}},
  \bibinfo{pages}{203} (\bibinfo{year}{1981}).

\bibitem[{\citenamefont{Kraichnan}(1967)}]{67Kra}
\bibinfo{author}{\bibfnamefont{R.~H.} \bibnamefont{Kraichnan}},
  \bibinfo{journal}{Phys. Fluids} \textbf{\bibinfo{volume}{10}},
  \bibinfo{pages}{1417} (\bibinfo{year}{1967}).

\bibitem[{\citenamefont{Connaughton et~al.}(2004)\citenamefont{Connaughton,
  Rajesh, and Zaboronski}}]{CRZtmshort}
\bibinfo{author}{\bibfnamefont{C.}~\bibnamefont{Connaughton}},
  \bibinfo{author}{\bibfnamefont{R.}~\bibnamefont{Rajesh}}, \bibnamefont{and}
  \bibinfo{author}{\bibfnamefont{O.}~\bibnamefont{Zaboronski}},
  \bibinfo{journal}{Phys. Rev. E} \textbf{\bibinfo{volume}{69}},
  \bibinfo{pages}{061114} (\bibinfo{year}{2004}).

\bibitem[{\citenamefont{Rajesh and Majumdar}(2000)}]{RM2000}
\bibinfo{author}{\bibfnamefont{R.}~\bibnamefont{Rajesh}} \bibnamefont{and}
  \bibinfo{author}{\bibfnamefont{S.~N.} \bibnamefont{Majumdar}},
  \bibinfo{journal}{Phys. Rev. E} \textbf{\bibinfo{volume}{62}},
  \bibinfo{pages}{3186} (\bibinfo{year}{2000}).

\end{thebibliography}

\end{document}